\documentclass[10pt,letterpaper]{article}
\usepackage[top=0.85in,left=2.75in,footskip=0.75in]{geometry}

\usepackage{amsmath,amssymb}

\usepackage{changepage}

\usepackage[utf8x]{inputenc}

\usepackage{textcomp,marvosym}

\usepackage{cite}

\usepackage{nameref}
\usepackage[hidelinks]{hyperref}
\hypersetup{colorlinks,linkcolor={magenta},citecolor={blue},urlcolor={magenta}} 

\usepackage[right]{lineno}

\usepackage{microtype}
\DisableLigatures[f]{encoding = *, family = * }

\usepackage[table]{xcolor}

\usepackage{array}

\usepackage{bm}

\newcolumntype{+}{!{\vrule width 2pt}}

\newlength\savedwidth

\usepackage[aboveskip=1pt,labelfont=bf,labelsep=period,singlelinecheck=off]{caption}

\makeatletter
\renewcommand{\@biblabel}[1]{\quad#1.}
\makeatother

\usepackage{lastpage,fancyhdr,graphicx}
\usepackage{epstopdf}
\fancyheadoffset[L]{2.25in}
\fancyfootoffset[L]{2.25in}
\newcommand{\LD}{\textcolor{black}}

\newcommand{\X}{\mathcal{X}}
\newcommand{\R}{\mathbb{R}}
\newcommand{\ie}{\emph{i.e.~}}

\begin{document}
\vspace*{0.2in}

\begin{flushleft}
{\Large
\textbf\newline{Estimating household contact matrices structure\\ from easily collectable metadata} 
}
\newline
\\
Lorenzo Dall'Amico\textsuperscript{1,*}
Jackie Kleynhans\textsuperscript{2,3}
Laetitia Gauvin\textsuperscript{1,4}
Michele Tizzoni\textsuperscript{1,5}
Laura Ozella\textsuperscript{1}
Mvuyo Makhasi\textsuperscript{2}
Nicole Wolter\textsuperscript{2,6}
Brigitte Language\textsuperscript{7}
Ryan G. Wagner\textsuperscript{8}
Cheryl Cohen\textsuperscript{2,3}
Stefano Tempia\textsuperscript{2,3}
Ciro Cattuto\textsuperscript{1,9}
\\
\bigskip
\textbf{1} ISI Foundation, Turin, 10126, Italy \\
\textbf{2} National Institute for Communicable Diseases of the National Health Laboratory Service, Johannesburg, South Africa \\
\textbf{3} School of Public Health, Faculty of Health Sciences, University of the Witwatersrand, Johannesburg, South Africa \\
\textbf{4} Institute for Research on sustainable Development, UMR215 PRODIG, Aubervilliers, France \\
\textbf{5} Department of Sociology and Social Research, University of Trento, Trento, Italy \\
\textbf{6} School of Pathology, University of the Witwatersrand, Johannesburg, South Africa \\
\textbf{7} Unit for Environmental Science and Management, Climatology Research Group, North-West University, Potchefstroom, South Africa \\
\textbf{8} MRC/Wits Rural Public Health and Health Transitions Research Unit (Agincourt) \\
\textbf{9} University of Turin, Department of Informatics, Turin, 10124, Italy \\
\textbf{*} Corresponding to: \tt{lorenzo.dallamico@isi.it}  

\bigskip

\end{flushleft}
\section*{Abstract}

\LD{Contact matrices are a commonly adopted data representation, used to develop compartmental models for epidemic spreading, accounting for the contact heterogeneities across age groups.} \LD{Their estimation, however, is generally time and effort consuming and model-driven strategies to quantify the contacts are often needed.}
\LD{In this article we focus on household contact matrices, describing the contacts among the members of a family and develop a parametric model to describe them. This model combines demographic and easily quantifiable survey-based data} and is tested on high resolution proximity data collected in two \LD{sites} in South Africa. Given its simplicity and interpretability, we expect our method to be easily applied to other contexts as well and we identify relevant questions that need to be addressed during the data collection procedure.



\section{Introduction}
\label{sec:intro}

Infectious diseases such as COVID-19 and influenza are transmitted through close proximity contacts \cite{wallinga1999perspective} and the modeling thereof is a problem of great interest for public health. The design of effective non-pharmaceutical interventions to mitigate the epidemic spreading often relies on models capable to predict the future or to reconstruct the past of the epidemic's state, see for instance \cite{anderson1992infectious, meyers2007contact, verity2020estimates, walker2020impact, sun2021transmission}. Households represent the minimal unit of disease transmission and play a fundamental role in determining the evolution of a viral spread \cite{house2009household}. Empirical evidences suggest that, especially at the household level, the commonly adopted homogeneous mixing hypothesis is insufficient to faithfully explain contagion \cite{goeyvaerts2018household, mccarthy2020quantifying, cencetti2021digital}. On the contrary, it is necessary to account for age-dependent \emph{contact matrices} that represent the diversities 	-- across different age classes -- in the frequency of contacts as well as in the transmission parameters \cite{wallinga2006using, hilton2019incorporating, li2020characteristics, edmunds1997mixes, mossong2008social}.

\medskip

Contact matrices are generally estimated through surveys in which the participants have to self-report their contacts in terms of number, duration and (presumed) age of the interacting individual \cite{mossong2008social, prem2017projecting, mistry2021inferring, potter2011estimating}. Known limitations of this technique include under-reporting of contacts and overestimation of their durations \cite{smieszek2012collecting, mastrandrea2016estimate}. Determining \emph{household contact matrices} (HCM) is resource-intensive, hardly scalable and technically challenging, especially in low-resource sub-Saharan African countries with high infectious diseases burden and where the data collection is still very limited \cite{johnstone2011social, kiti2014quantifying, de2018characteristics, thindwa2022social, naghavi2017global}. 	
\LD{Consequently, a growing attention is devoted to theoretically model HCM. Some of the most popular models to estimate contact matrices rely on the demographic properties of the population under study \cite{fumanelli2012inferring, mistry2021inferring}, eventually taking the setting (\emph{e.g.} school, work, home) in which the interactions take place into account. These models assume that the number of contacts between age groups approximately scales as the product of the two population sizes involved, \ie the number of all possible pairs. In \cite{prem2017projecting} the authors further considered how to 	 make estimates of contact matrices available in countries where the mixing patterns were not directly estimated. 
More recently, \cite{Manna2023GeneralizedCM} introduced generalized contact matrices in which socio-economic factors are included as well. The authors propose a simple model inducing assortative mixing 	that is pervasively observed in real-world data.}

\medskip

Here we consider HCM obtained from proximity sensors, encoding the sequence of contacts among a group of selected participants with high resolution in space and time. 
The proximity sensors are developed by the \texttt{SocioPatterns} collaboration (\href{http://www.sociopatterns.org/}{sociopatterns.org}, \cite{cattuto2010dynamics}) and allow us to  study and model human dynamics \cite{johnstone2011social, stehle2011high, vanhems2013estimating, ozella2018close, starnini2017robust, kiti2019study} and directly estimate HCM by aggregating individuals' contacts across time. We analyze the data collected during the PHIRST study \cite{cohen2021cohort, kleynhans2021cross}, a $3$-year long experiment conducted in South Africa, designed to provide reliable data-driven guidance to limit viral transmission \cite{cohen2021cohort, cohen2021sars, kleynhans2021sars, cohen2021asymptomatic, thindwa2021estimating, wilkinson2021year, igboh2021influenza, tempia2021decline}.  \LD{We show that, although demographic properties are determinant in shaping the HCM, they are insufficient to accurately capture the contacts structure and further age-dependent parameters must be introduced to model the higher sociability typically observed among young people \cite{hoang2019systematic}. Our parametric model can be calibrated with surveys but, unlike the direct estimation of the full contact matrix, they introduce several advantages. Firstly one only needs to report one's age and not the age of the other interacting individuals, making the estimation process more reliable by design. Secondly, the number of parameters to be estimated scales linearly with the number of age bins (and not quadratically) and the binning itself can be chosen a posteriori. Our method can thus be seen as a reliable compromise between a parameter-free demographic model and a direct estimation of the contact matrix from surveys. Testing our results on the high-resolution measurements, we show that one can approximate the HCM with a cosine similarity equal to $0.96$ and $0.98$ in the two sites.}

\begin{figure}[!t]
	\centering
	\includegraphics[width=\linewidth]{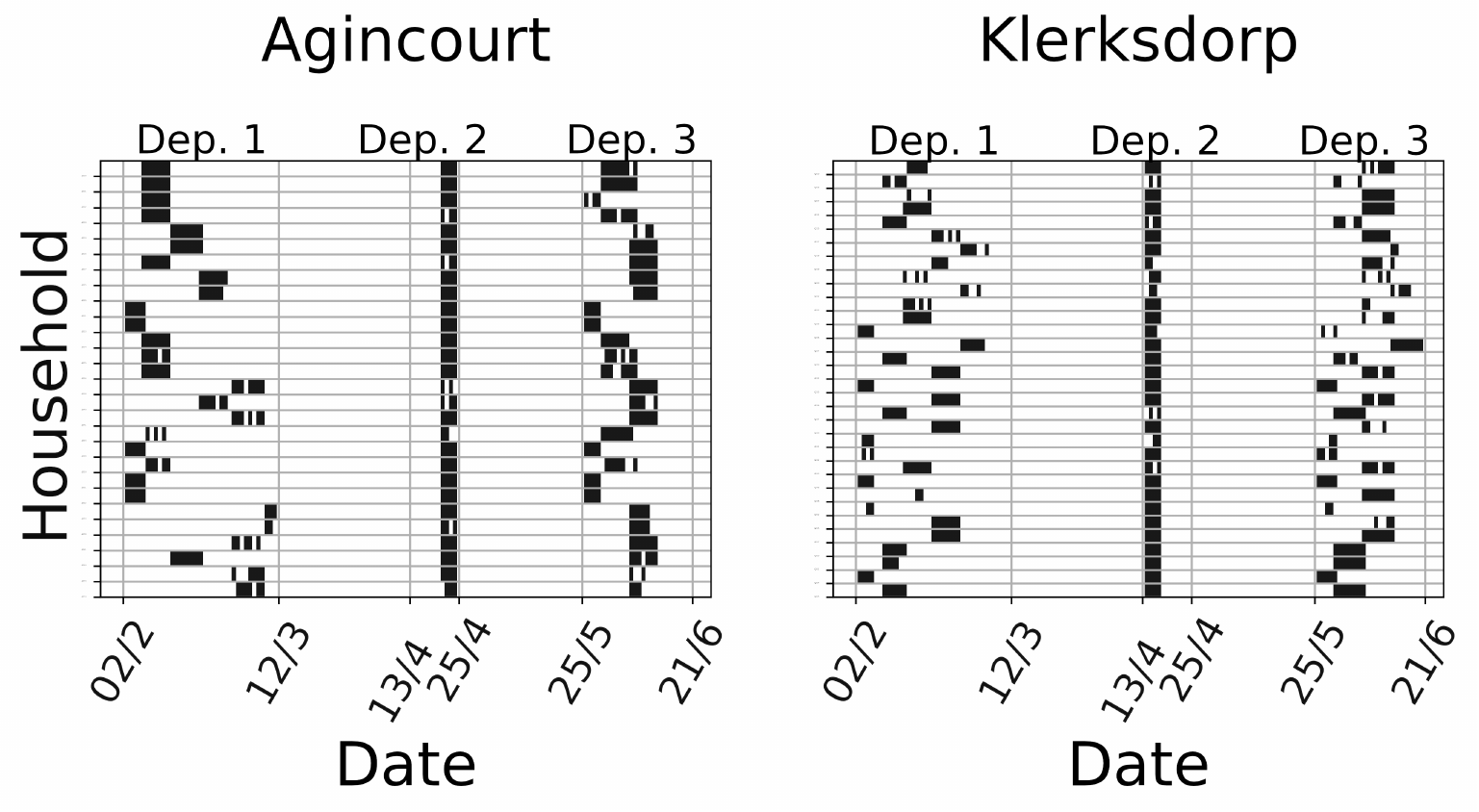}
	\caption{\textbf{Data collection schedule for the $60$ selected households}. Each row corresponds to a household \LD{with the rural site on the left and the urban site on the right}. Time is displayed on the $x$ axis and dates are reported in the day/month format. Vertical gray lines correspond to the beginning and end of each deployment. A black dot indicates that at least one contact was measured, while a white one that no contact was recorded on that day.}
	\label{fig:data_collection}
\end{figure}

\begin{figure}[!t]
	\centering
	\includegraphics[width=\linewidth]{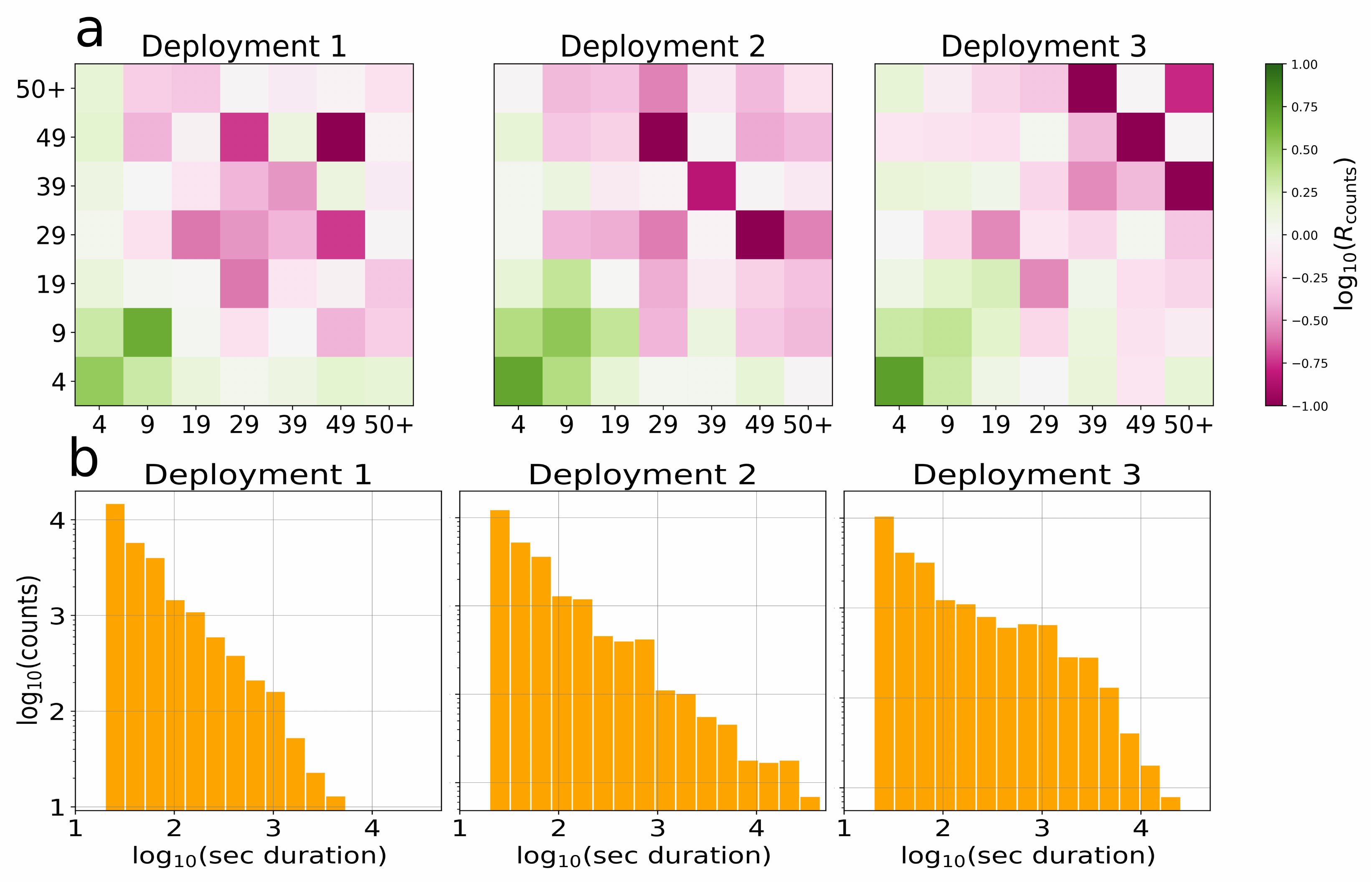}
	\caption{\textbf{Properties of the measured data}: \textbf{a}: normalized contact matrix across the three deployments. The color code refers to the values of the logarithm of $R_{\rm counts}$ whose entries are proportional to the ratio between the number of contacts and the number of possible interacting pairs, setting the mean of $R_{\rm counts}$ to 1. The two axis correspond to the age groups and the number reported indicates the highest age of each group. \textbf{b}: contact duration distribution \LD{expressed as number of seconds of interaction} across the three deployments in logarithmic scale.}
	\label{fig:CM}
\end{figure}

\section{Data descriptive statistics}
\label{sec:describe}

\LD{We now provide an overview of the data collection strategy, as well as some basic descriptive statistics.}

\subsection{Data collection}
\label{sec:describe.data}

The PHIRST study was a prospective household cohort study described previously in \cite{cohen2021cohort, cohen2021asymptomatic}. We enrolled a cohort of households $2018$ at two sites in South Africa (urban: Klerksdorp, North West and rural: Agincourt, Mpumalanga) and followed households up for $8$ to $10$ months.  Wearable proximity sensors were deployed for $10$ to $14$ days to all consenting household members to measure high-resolution household contact patterns during three periods of the year. Sensors were worn in PVC pouches on the chest or on a lanyard. Participants were requested to wear the sensor on in the morning, keep it on the entire day (even when leaving the home), take it off at night and store it separately from other household member’s sensors. Not all participants felt comfortable wearing sensors outside of the home and instead took sensors off when not at home. Participants were requested to complete a diary to indicate the times the sensor was put on and taken off during the day. Twice a week, the staff visited each household and reminded participants to wear the sensors, monitored if all sensors were still working, and replaced batteries where sensors had stopped working. After at least a ten-day deployment, sensors were collected at the next routine household visit of study staff to the household and taken back to the study office where batteries were removed \LD{and data was downloaded from the sensors}. \LD{After the data cleaning procedure, detailed in Section~\ref{app:data}, our dataset is composed of $307$ individuals subdivided into $60$ households. For consistency, we choose to consider only households for which the data quality was sufficiently high in all three deployments. The exclusion can be due to the displacement of some individuals or to  technical problems with specific sensors. As discussed in the supplementary material, the cleaned dataset is representative of the original both in terms of size and age distributions. Figure~\ref{fig:data_collection} summarizes the data collection schedule.}

\subsubsection*{Ethical approval}

All experiments were performed in accordance with relevant guidelines and regulations: ethics permission to conduct the experiment was received from the Wits Human Research Ethics Committee (Medical) (ethics reference no. 150808) as well as the Mpumalanga Provincial Research \& Ethics Committee. Written informed consent was sought and received from all participants or their caregivers.

\subsection{Contact matrices}
\label{sec:describe.CM}

\LD{In this section we describe the properties of the contact matrices as measured by the proximity sensors, after having provided some formal definitions.}

\subsubsection*{Definitions}

\LD{Contact matrices incorporate the contacts subdivided by age groups.} They are square and symmetric, of size $n_{\rm age}$, the number of age bins considered. Here the age groups are divided into $[0-4,5-9,10-19,20-29,30-39,40-49,50+]$ years: the finer grain of younger ages is because of the large proportion of population in those age brackets, shown in Figure~\ref{fig:supp_data}. \LD{Each HCM refers to a single household and a specific deployment. We thus consider a total of $180$ HCM.}
With the notation $C, S$ we refer to the contact matrices storing the counts/time of interaction between pairs of age groups respectively, or, more precisely

\begin{align*}
C_{ab} &= {\rm~ number~of~contacts~per~day~between~} a~{\rm and~}b, \\
S_{ab} &= {\rm~ total~time~in~contact~per~day~between~} a~{\rm and~}b.
\end{align*}

These matrices should be compared with their expectation, \ie with the contact matrix obtained assuming a given household line-up and that people interact at random. This is given by \cite{fumanelli2012inferring}:

\begin{align}
T_{ab} = \frac{\Phi_a\Phi_b - \delta_{ab}}{\rho-1},
\label{eq:T}
\end{align}

where $\Phi_a$ is the number of people in the age group $a$ in a given HCM; $\rho = \sum_a \Phi_a$ is the total number of people and \LD{$\delta_{ab}$ is the Kroeneker delta (equal to $1$ is $a=b$ and equal to $0$ otherwise). For a set $\X$ of HCM, we define $C^{(\X)}, S^{(\X)}, T^{(\X)}$ as the average of the respective matrix over all $\X$ and $R_{C}^{(\X)}$ as}

\begin{align*}
\left(R_{C}^{(\X)}\right)_{ab} = 
\begin{cases}
\gamma^{(\X)}~\frac{C^{(\X)}_{ab}}{T_{ab}^{(\X)}}~\quad &{\rm if~} T_{ab}^{(\X)} \neq 0\\
1 \quad &{\rm else}
\end{cases}
\end{align*}

where $\gamma^{(\X)}$ is a constant to impose that the average of $R_C^{(\X)}$ equals one. \LD{In an analogous way, we define $R_S^{(\X)}$ replacing $C$ with $S$. In words, the entries of $R^{(\X)}$ exceed one for the pairs that interact more than expected and are below one otherwise. If a pair cannot have interactions, we conventionally set $R^{(\X)} = 1$. To simplify the notation, in the remainder we drop the index $\X$.}

\subsubsection*{Properties of the measured matrices}

\begin{table*}[!t]
	\def\arraystretch{1.2}
	\centering
	\begin{tabular}{|c || c c c|} 

		\hline

		$R_{C}$& First & Second & Third \\ [0.5ex] 

		\hline\hline

		First & 1 & 0.94 & 0.89 \\

		Second& 0.94 & 1 & 0.94 \\ 

		Third & 0.89 & 0.94 & 1\\

		\hline

	\end{tabular}
	\quad
	\begin{tabular}{|c || c c c|} 

		\hline

		$R_{S}$& First & Second & Third \\ [0.5ex] 

		\hline\hline

		First & 1 & 0.85 & 0.87 \\

		Second& 0.85 & 1 & 0.87 \\ 

		Third & 0.87 & 0.87 & 1\\

		\hline

	\end{tabular}

	\caption{\textbf{Contact matrix similarity across the deployments}. Cosine similarity between the measured contact matrices $R_{C}$ (left) and $R_{S}$ (right) in the three deployments.}

\label{table:main.sim}

\end{table*}

\LD{Given that we considered the same set of households across the three deployments, changes in the HCM structure can mainly be amenable to a seasonality effect}. Table~\ref{table:main.sim} precisely shows the cosine similarity between $R_{C}$ (left) and $R_{S}$ (right) for $\mathcal{X}_1, \mathcal{X}_2, \mathcal{X}_3$ being the set of all households in the three deployments.  The table reports high similarity values for $R_{C}$, suggesting that the structure of the contact matrix does not vary a lot across the three deployments. Smaller values are instead obtained by $R_{S}$ implying that the seasonality effect majorly involves the duration (rather than the structure) of the contacts. This observation agrees with the distribution of the \LD{ individual contact durations, obtained from approximately $10^5$ proximity measurements} shown in Figure~\ref{fig:CM}b which follows \LD{a broad distribution, as expected} \cite{barabasi2005origin}. This distribution broadens in the third deployment when south-African winter is approaching. More quantitatively, we computed the $99^{\rm th}$ percentile for the three distributions that is approximately $12$ minutes in the first deployment, $27$ in the second and $60$ in the last. Figure~\ref{fig:CM}a shows instead the matrix ${\rm log}(R_{C})$ across the three deployments, evidencing that younger age groups tend to interact more, regardless of the age group they are interacting with. Based on these observations, we attempt to model the matrix $C$ whose behavior is more predictable than $S$. Given the result of Table~\ref{table:main.sim}, the deployments are treated as three independent, equally reliable measurements of the HCMs.

\section{Main result}
\label{sec:results}

We introduce two parametric models to approximate the HCM  that combine three age-dependent parameters: the number of individuals per age group, \LD{the in-house hourly presence} and an intensity of activity factor. All the parameters involved in the model only depend on a single age class and not on the interactions between pairs of age classes, as it is commonly required in self-reporting surveys. This allows us to decrease the number of parameters to be estimated from order of $n_{\rm age}^2$ to $n_{\rm age}$.

\medskip

We here propose some example of questions to estimate the \LD{in-house hourly presence and the intensity of activity factor.}

	\begin{enumerate}

		\item[$\bullet$] \emph{How much time do you typically spend at home in each hour of the day?}

		\item[$\bullet$] \emph{How much of this time do you typically spend in isolation?}

		\item[$\bullet$] \LD{\emph{How many face-to-face interactions do you have per day?}}

	\end{enumerate}

As we will see in the remainder, these questions permit to calibrate the parameters of our model, allowing one to obtain a more faithful representation of contact matrices than the one obtained from purely demographic models. \LD{In Section~\ref{sec:main.practical} we describe some practical implications of our results and the relation to the questions listed above.}

\begin{figure}[!t]
	\centering
	\includegraphics[width=\linewidth]{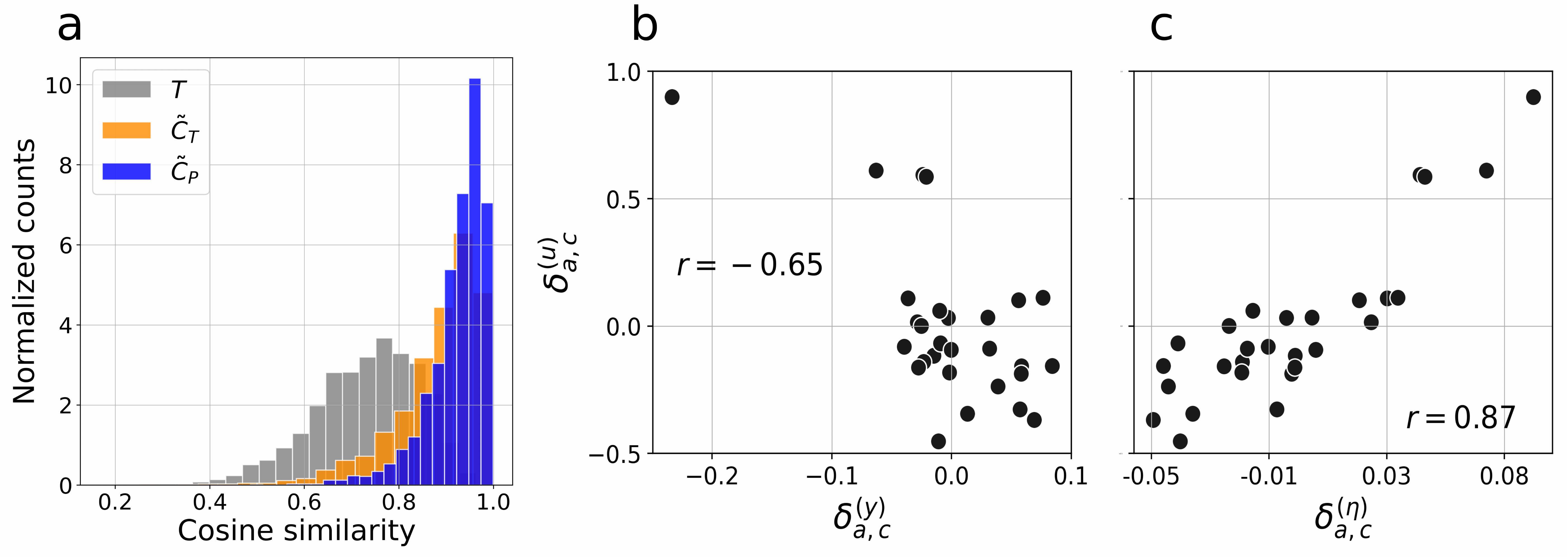}
	\caption{\textbf{Test of the model for household interaction}. \textbf{a}: histogram of the cosine similarity between $C$ and its estimators. The gray curve corresponds to the histogram over the $2500$ realization of $\X$ using $T$ as an estimator of $C$.	The orange curve is obtained with the first order model of \eqref{eq:CT}, while the blue curve corresponds to the second order model of Section \ref{sec:main.second}. \textbf{b}, \textbf{c}: correlation between the fluctuations of the activity $\delta^{(u)}$, the group average degree $\delta^{(\eta)}$ and the presence of a major occupation outside the house $\delta^{(y)}$. The quantities $\delta_{a,c}$ are defined in Equation \eqref{eq:main.fluct}. The Pearson correlation coefficient $r$ is reported in text.}
	\label{fig:res}
\end{figure}

\subsection{A first order model for household interaction}

\label{sec:main.first}

\LD{In this section, we define a parametric model to approximate the contact matrix $C$, as measured by proximity sensors. All matrices here refer to sets of HCM but we drop the index $\X$ to keep a light notation. Let $T$ be the matrix defined in Equation~\eqref{eq:T}.
We define $\tilde{C}_{T}$, an approximation of $C$, as
\begin{align}
\label{eq:CT}
\tilde{C}_{T} = T \circ \left(\bm{u}\bm{u}^T\right),
\end{align}}

where $\bm{u}\in \R^{n_{\rm age}}$ is a set of parameters that represent the activity of each age group and `$\circ$' denotes the entry-wise Hadamard product. \LD{The entries of this matrix are $(\tilde{C}_{T})_{ab} = T_{ab}u_au_b$} and a large number of interactions are expected when many members are present (large values of $T_{ab}$) and when they correspond to highly active age groups, such as $[0 - 4, 5 - 9]$, as per Figure~\ref{fig:CM}b.

\subsubsection*{Model validation}

We deploy the following steps to test our model, as detailed and motivated in Section~\ref{supp:method}. \LD{We independently randomly sample $2500$ sets $\X$ of $8$ HCM without replacement out of the $180$ available}. For each sampled $\X$ we compute the vector $\bm{u}$ that best approximates $C$, \LD{minimizing a modified Canberra distance \cite{lance1966computer} between the measured and the estimated matrix, as described in Section~\ref{supp:method}}. The entries of this vector contain the activity of each age group for the set $\X$. Figure~\ref{fig:res}a displays the histogram of the cosine similarity between the approximation $\tilde{C}_{T}$ and the measured matrix $C$ and evidences a good agreement between the two matrices with a cosine similarity equal to $0.9$ or larger for $53\%$ of the data. \LD{This similarity is of the same order of the one observed across the three deployments and reported in Table~\ref{table:main.sim}}. Figure~\ref{fig:res} further shows the same histogram for $T$ being used as an estimator of $C$. This purely demographic model is much less accurate and reaches a cosine similarity greater than $0.9$ for only $7\%$ of the data and \LD{$50\%$ of the data have a similarity greater or equal to $0.75$.}

\subsubsection*{Interpretation of the parameters}

Besides the goodness of the approximation itself, our main interest is to assess whether the vector $\bm{u}$ can be estimated from easily observable quantities. To do so, for each sampled $\X$ we further compute the vector $\bm{\eta} \in \R^{n_{\rm age}}$. \LD{Its element $\eta_a$ is the number of daily interactions per individual, averaged over all individuals in a given age group $a$.}
Intuitively, $\bm{u}$ and  $\bm{\eta}$ should correlate: a higher activity has to be observed when people are more active. Note that $\bm{\eta}$ aggregates \emph{all} individual's contacts and is oblivious of the age group binning. We divide the sets $\X$ according to their activity vector representation $\bm{u}$ into $k = 4$ groups with a hierarchical clustering algorithm.
For each	 $\bm{x} \in \{\bm{u}, \bm{\eta}\}$, we then write the value corresponding to age $a$ and class $c$ as

\begin{align}
x_{a,c} = \bar{x}_a + \delta^{(x)}_{a,c},
\label{eq:main.fluct}
\end{align}

where $\bar{x}_a$ is the average over the $4$ groups, and $\delta^{(x)}_{a,p}$ are the fluctuations. Figures~\ref{fig:res}b shows the scatter plot of the fluctuations of $\bm{\delta}^{(u)}$ and $\bm{\delta}^{(\eta)}$, evidencing a strong correlation with a highly significant ($p$-value less than $10^{-3}$) Pearson coefficient of $0.85$.

\medskip

This analysis suggests that the measured contact matrix can be estimated with a high precision from aggregated (hence more easily collectable) data being the average number of contacts per individual \LD{in the same age group}. We now introduce a further parameter $\bm{y}$ that is even more easily observable than $\bm{\eta}$ and has a weaker but still strong correlation with $\bm{u}$. Specifically, the entries of $\bm{y} \in [0,1]^{n_{\rm age}}$ indicate the fraction of people for each age group having an occupation outside the house \LD{requiring at least three hours a day}. This quantity is expected to be negatively correlated with $\bm{u}$, since lower activities should be observed when people spend more time outside the household. Repeating the same procedure detailed for $\bm{\eta}$, we obtain Figures~\ref{fig:res}f showing indeed that high values of $u_a$ are obtained for low $y_a$, as expected (see the red squares for the group $[40 - 49]$). The correlation between the fluctuations of $\bm{u}$ and $\bm{y}$ is reported in Figure~\ref{fig:res}c, reaching a significant Pearson coefficient of $-0.65$.
We underline that $\bm{y}$ is a very aggregated quantity that does not directly involve contacts.

\medskip

We now discuss a refined model with respect to Eq~\eqref{eq:CT} that keeps simultaneously into account the activity and the time spent at home. We show that this model produces better estimates  of the contact matrices and can be conveniently used to predict the HCM originally excluded from our study.

\begin{figure}
	\centering
	\includegraphics[width=\columnwidth]{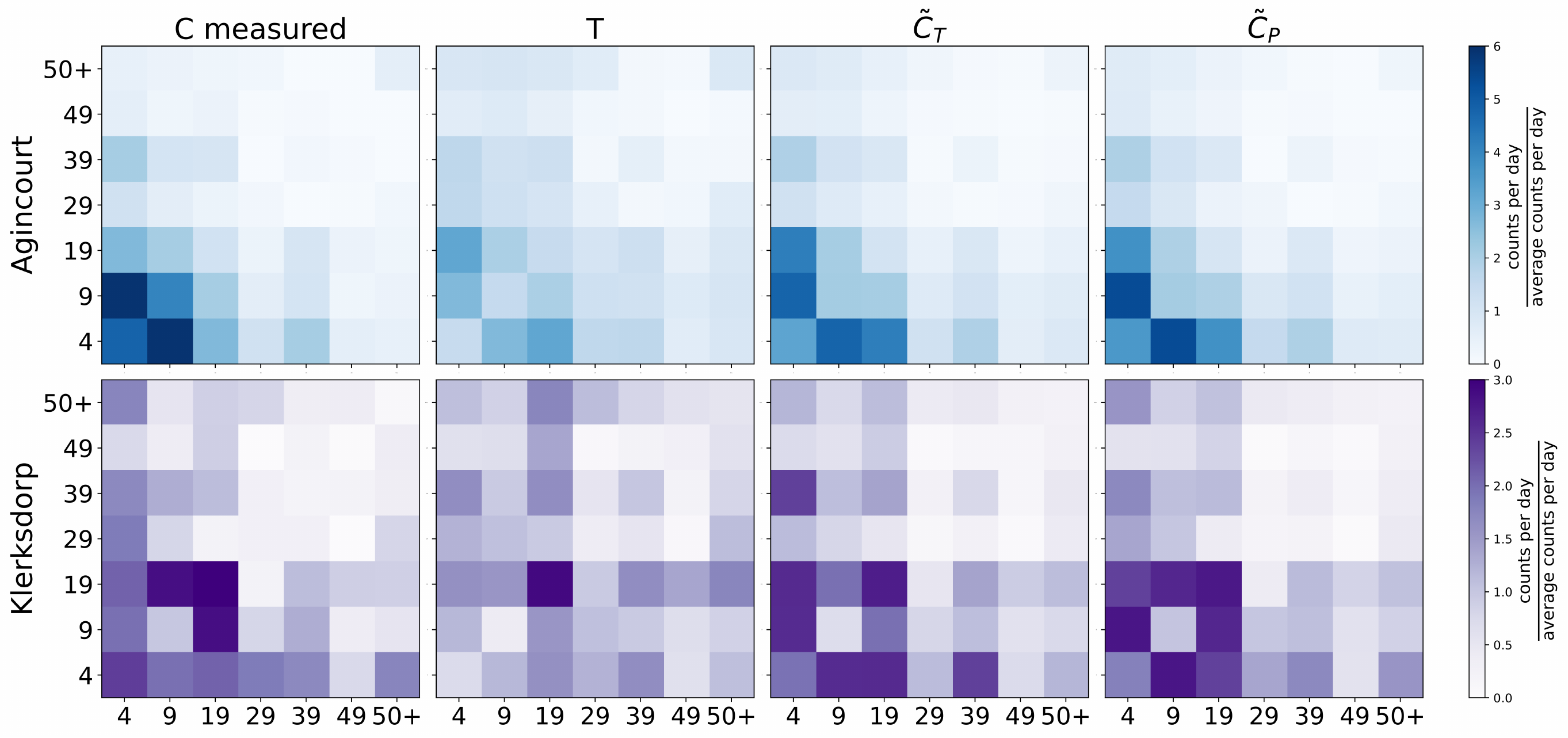}
	\caption{\textbf{Measured vs estimated normalized contact matrices in the two sites}. The first row, in blue, corresponds to Agincourt, the rural site, while the second, in purple, to Klerksdorp, the urban site. The first column shows the matrix $C$ aggregated over the three deployments, as measured by the proximity sensors. The second column is the corresponding random encounter matrix $T$. The third and the fourth are the estimates obtained by our first and second order models, respectively. All matrices are normalized by the empirical average of their entries.}
	\label{fig:est}
\end{figure}

\subsection{A second order model for household interaction}

\label{sec:main.second}

\LD{In Equation \eqref{eq:T} we introduced the matrix $T$ that encodes a purely demographic interaction model in which a higher contact rate is entirely explained by a higher number of interacting individuals. In practice, however, contacts can happen only when people are in the same physical space. To model this effect, we propose an extension of $T$, that we denote with $P$. Let $\bm{v}_i \in \{0,1\}^{24}$ be a binary-value presence vector of $i$, denoting the presence in the house for each hour of the day. The definition of $P$ then reads}

\begin{align}
P_{ab} = \frac{1}{\rho-1}\sum_{i \in V_a}\sum_{j \in V_b\setminus \{i\}}\frac{\bm{v}_i^T\bm{v}_j}{24}
\label{eq:P}
\end{align}

where $V_a$ is the set of all individuals in the age group $a$. Note that if $\bm{v}_{i,t} = 1$ for all $i$ and all $t$, the definition of $P$ corresponds to the one of $T$. The scalar product between $\bm{v}_i^T\bm{v}_j$ quantifies the time in which $i$ and $j$ had \emph{simultaneously} contacts with members inside the household. If it equals zero, then there is no chance that $i$ and $j$ got in contact at all. In other words, $P$ predicts the contact rate assuming people get in proximity at random, but keeping into account that people are not always and simultaneously inside the house. We generalize the model of Eq~\eqref{eq:CT} replacing $T$ with $P$ and obtaining $\tilde{C}_P$.  \LD{Practically, the proximity sensors do not provide us with the information of whether or not an individual is at home in a given moment, but only if it is interacting with another household member. For each individual we then construct a binary indicator on whether or not he/she interacted with someone in a particular hour of the day during the deployment and use this as a proxy for $\bm{v}$.}

\subsubsection*{Model testing}

The blue histogram of Figure~\ref{fig:res}a shows the cosine similarity between the actual and estimated contact matrices obtained using $P$.  A clear gain in accuracy is achieved, obtaining a cosine similarity is greater than $0.9$ for $75\%$ of the data.

\medskip

We finally test the goodness of our model for the two sites separately on all \emph{(household-deployment)} valid pairs, hence also those that were initially excluded because of quality issues in some (but not all) deployments. We use as $\bm{u}$ its average realization over the $2500$ samples and compare the result of the predicted matrix $T, \tilde{C}_T$ and $\tilde{C}_{P}$ with the measured one (Figure~\ref{fig:est}), considering the two sites separately. The cosine similarity scores reported in Table~\ref{tab:cos} provide and striking evidence of how contact matrices are approximated with high precision using few age-dependent parameters.

\begin{table}[t!]

	\centering

	\bgroup

	\def\arraystretch{1.4}

	\begin{tabular}{|c || c c c|} 

		\hline

		& $T$ & $\tilde{C}_T$ & $\tilde{C}_P$ \\ [.5ex] 

		\hline\hline

		Agincourt & 0.83 & 0.95 & 0.96 \\

		Klerksdorp& 0.89 & 0.95 & 0.98 \\ 

		\hline

	\end{tabular}

	\egroup

	\caption{\textbf{Goodness of the contact matrix estimation for different methods}. The score is reported in terms of cosine similarity and the naming is consistent with Figure~\ref{fig:est} which this table refers to.}

	\label{tab:cos}

\end{table}

\subsection{Practical implications}
\label{sec:main.practical}

\LD{Let us briefly discuss some implications of our results and suggest how these could be translated into practical recommendations for data collection. Survey based estimations are, to-date, the most common and reliable way to estimate contact matrices. This method, however, has some notable limitations -- that we discussed in the Section \ref{sec:intro} -- and would  benefit from the design of simpler questionnaires. We highlight that one can accurately estimate HCM from self-reported quantities that are, by design, more easily and reliably estimated. Our model combines the probability that two individuals meet with an age-dependent activity driven model \cite{perra2012activity}.}

\medskip

\LD{We suggested some examples of questions that can be formulated to calibrate our model. For instance, the question ``\emph{How much time do you typically spend at home in each hour of the day?}'', can be used to quantify the vectors $\bm{v}$ of Equation \eqref{eq:P}, needed to obtain $P$. The similarity of these vectors gives already a good estimation of the probability of interaction of the household members. Even if our experiment focused only on the household contacts, we envision that this approach can be directly extended to other settings, designing context-related contact matrices as done in \cite{fumanelli2012inferring}. Moreover, one can think of providing a finer estimation of $\bm{v}$ considering a multi-day average, so that $v_t \in [0,1]$ is a probability to be at home (or, more generally, in a given place) at time $t$. The question ``\emph{How much of this time do you typically spend in isolation?}'' then can allow one to re-weight the entries $v_t$ to account for an actual probability of encounter. The last question ``\emph{How many face-to-face interactions you have per day?}'' is an example of how one can quantify an individuals' activity rate. Given these estimates, the age parameters are obtained simply aggregating them according to the relevant age-group to obtain the activity vector $\bm{u}$.}   

\section{Conclusion}
\label{sec:discussion}

Our result brings an empirical evidence that most of the structure of contact matrices measured with high-resolution proximity sensors can be reliably captured with a simple statistical model \LD{combining behavioral parameters with demographic ones}.
While it comes as no surprise that a generalization of the matrix $T$ would lead to better estimates, the most important aspects of our results are listed as follows:

\begin{itemize}

	\item Simple, environment-independent models can accurately estimate HCM. The high quality and size of the PHIRST dataset gave us great insights into the problem of HCM estimation. Backed by these empirical data, not only can we say that the proposed parametric model generally improves the estimation accuracy, but we can numerically quantify it, observing very high level of agreement with the HCM obtained with the costly high resolution measurements
	\item Our proposed models are highly interpretable. We expect its parameters to be  easily estimated with surveys, addressing questions such as those listed in Section~\ref{sec:intro}. We expect this to be one of the significant outcomes of our research as we identified some practical questions to calibrate our model, bypassing proximity sensors. 

	\item All parameters are aggregated by age group and involve the behavior of single individuals and do not depend on the age class of other members. This aspect naturally reduces the number of parameters of the model, making the estimation process simpler and addresses the important requirement for surveys that the questions asked should have a simple answer.

\end{itemize}



The questions suggested in Section~\ref{sec:intro} constitute an example of possible ways to estimate the activity parameters and are limited to the quantities that turned out to provide a significant explanation of HCM in our experiment setting. Other metadata  (such as the number of rooms in the house, the wealth status or the distinction between the rural and the urban site) could potentially be informative to explain the HCM structure, even if they were not in our analysis. 

%
%

\medskip

The main limitations of our methodology are related to the quality and nature of the available data. The first concern is related to the time-dependent data collection component which we essentially neglected here. When dealing with contact matrices, it is customary to distinguish between weekdays and weekends. In our measurements, the first and third waves of measurements in households were made asynchronously. After the cleaning procedure, it emerged that, \LD{as a consequence of the adoption of this choice for the scheduling of data collection in the field}, weekdays and weekends are not evenly distributed among households and changes in the measured HCM are potentially associated with this effect. To cope with this problem, when dealing with asynchronous measurements it would be preferable to consider the same days of the week for all households. A closely related concern is that we have considered all three deployments as equal, even though they correspond to rather different periods in the year. The data sparsity and quality did not allow us to detect any significant change \LD{in the seasonality of the contact patterns}, except for the duration of contact distribution shown in Figure~\ref{fig:CM}c. It is nonetheless a very reasonable assumption that the contact behavior changes during the year. Our suggestion to investigate individuals' behavioral habits can easily overcome this problem, designing time-dependent expected matrices that could adapt even to diverse scenarios such as, during a quarantine.

\medskip

\LD{In conclusion, our study proposes a parametric model to estimate contact matrices with high accuracy. It improves over the purely demographic models in terms of accuracy and over the purely survey-based approaches in terms of simplicity of the data collection. Given its simplicity and interpretability, we envision that our framework can be adopted to estimate contact matrices beyond the household setting.}
As a practical application, our results can impact the strategy to design the surveys currently adopted to quantify social contacts to mitigate the Covid19 and similar epidemics \cite{verelst2021socrates, koppeschaar2017influenzanet}.

\section*{Data availability}

The contact matrices aggregated at the household level are made available at\\ \href{https://github.com/lorenzodallamico/PHIRST\_CM/}{github.com/lorenzodallamico/PHIRST\_CM}.


\section*{Funding}

This work was supported by the National Institute for Communicable Diseases of the National Health Laboratory Service and the U.S. Centers for Disease Control and Prevention [co-operative agreement number: 1U01IP001048]. The funders had no role in design, analysis or interpretation of data.
LD and CCa acknowledge support from the Lagrange Project of ISI Foundation funded by CRT Foundation, from the European Union’s Horizon 2020 research and innovation programme under grant agreement No. 101016233 (PERISCOPE) and from Fondation Botnar.
LG, MT and LO acknowledge support from the Lagrange Project of ISI Foundation funded by CRT Foundation.

\section*{Competing interests}

CCo has received grant support from Sanofi Pasteur, US CDC, Welcome Trust, Programme for Applied Technologies in Health (PATH), Bill \& Melinda Gates Foundation and South African Medical Research Council (SA-MRC). NW reports receiving grants from Sanofi Pasteur, US CDC and the Bill \& Melinda Gates Foundation. All other authors do not report any competing interests.

\nolinenumbers

\newpage

\renewcommand{\thesection}{S}
\setcounter{subsection}{0}
\setcounter{figure}{0}
\renewcommand{\thefigure}{S.\arabic{figure}}

\normalsize


\section{Supplementary information: Appendix}

\subsection{Data collection and pre-processing}
\label{app:data}

\begin{figure}[!t]
	\centering
	\includegraphics[width=\columnwidth]{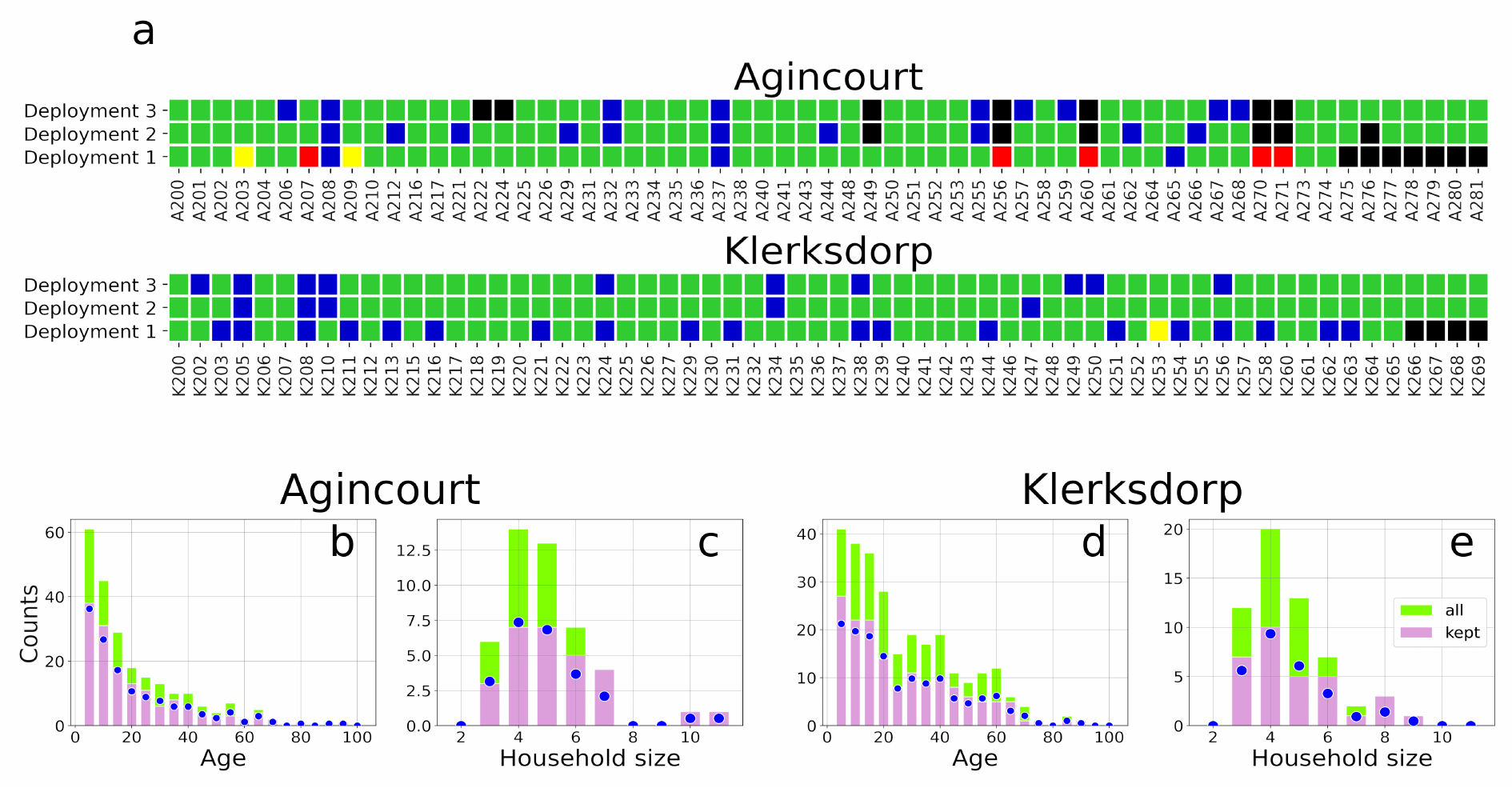}
	\caption{\textbf{Raw data characteristics}. \textbf{a}: data quality. \LD{On the x-axis we plot households, while on the $y$-axis the deployments.} For each (household-deployment) we assign a color code: black indicates that the household did not participate; red that all household’s sensors had data quality issues and did not provide valid measurements; blue that there are less than two days of measurement; yellow that a non circadian activity is observed; green none of the above. \textbf{b} and \textbf{d}: age distribution in Agincourt and Klerksdorp, respectively. The green bars are referred to the whole data-set, while the purple one only refers to the $60$ households with valid measurements in all three deployments (see \textbf{a}). Blue dots are obtained by multiplying the height of the green bars for the fraction of the included households, \LD{that is the expected bar height, given the cleaned dataset size}. \textbf{c} and \textbf{e}: household size distribution. Legends and colors follow \textbf{b} and \textbf{d}.}
	\label{fig:supp_data}
\end{figure}

Proximity data are measured with the \texttt{SocioPatterns} sensors that we here introduce, addressing the interested reader to \cite{cattuto2010dynamics} for a more detailed reference. Their functioning is based on the emission of low-power signals. Participants are asked to wear the sensor on their chest, so that when they engage in a face-to-face interaction with another participant, the respective sensors can exchange packets of information with a frequency that does not exceed one packet per second. A contact is measured if, in the time-span of $20$ seconds, two sensors exchange at least one packet, recording the unique identifier of the interacting sensor, the time at which the interaction occurred and the attenuation of the signal from the sender to the receiver. This attenuation is related to the distance between the two sensors and can be used to filter suitably-defined close-range proximity relations. \LD{Additionally, each sensor periodically records some status properties that log metadata and diagnostic information}. Among these, an accelerometer allows one to know \LD{every $15$ minutes} if the sensor is moving or not. \LD{Given the sensitivity of the accelerometer and the time-scale at which it operates, one can assume that if the sensor is still, then it is not worn.} The cleaning procedure is summarized as follows:


\begin{enumerate}
	\item All contacts measured by non-moving sensors are removed: this is to avoid including spurious contacts between sensors that are, for instance, kept inside a drawer
	\item Contacts are filtered  and only those with a \LD{suitable attenuation threshold}. This threshold corresponds to an interaction between two sensors that are approximately at $2$ meters, even if this is a context-dependent relation that depends on external parameters, such as, for instance, humidity.
	\item All contacts happening before the beginning of the deployment (as reported in the diaries) and after its end are removed. These contacts may exist, because sensors may be collected on different dates from the ones of the planned experiment, but they are removed because sensors' use may be non-systematic, hence unreliable. Moreover, the first and last day of measurement are removed as well. During these days, very intense activity patterns are typically observed due to the interaction with the people dispatching the sensors. Since this kind of interaction deviates from the standard conditions, it is not considered.
	\item The data collected by the sensors contains information on the hardware identification code. A mapping relates this identifier with the individuals' pseudonym that allows us to relate contacts and metadata. Errors at this stage make it impossible to relate contacts to people and results in the red dots shown in Figure~\ref{fig:supp_data}a.
	\item As a minimal request, we impose that, after this cleaning procedure, a deployment can be considered valid only if it has two or more days of measurement. We found this to be a good trade-off between high quality data to work with and a sufficiently comprehensive inclusion principle. Household-deployment pairs that do not fulfill this condition are denoted in blue in Figure~\ref{fig:supp_data}a.
	\item \LD{Finally, non-circadian activity patterns are identified. A great excess of activity during night hours was observed in three households (yellow dots of Figure~\ref{fig:supp_data}a) during the first deployment. This may occur, for instance, if the sensors are left in proximity on a vibrating surface: the accelerometer filter does not remove these contacts even though the sensors were not worn at that moment.}
\end{enumerate}

Only the households in which all three deployments led to valid measurements (all green dots in Figure~\ref{fig:supp_data}a) were included in our study. Figure~\ref{fig:supp_data}b, c, d, e further show the age and household size histograms for the whole dataset against its cleaned version, showing that our inclusion principle did not affect either of the four distributions.

\subsection{Validation approach}
\label{supp:method}

\subsubsection*{Sampling the villages}

First of all, \LD{in order to devise a good approximation of HCM}, it is necessary to define a suitable distance to compare them. When comparing different households, however, one has to consider that typically there are some age groups with no individuals. More formally, this means that for some age group $a$, $\Phi_a = 0$. \LD{In some extreme cases there is no way to consistently compare HCM because the corresponding contact matrices are complementary, \ie the zeros one correspond to the non-zeros of the other}.

\medskip

To address this problem, we choose to compare groups of HCM (\emph{villages}) $\X$, \ie small groups of household-deployment $(h,d)$ pairs that guarantee that $\Phi_a > 0$ for all $a$. To build the samples $\X$ we then first select some $(h,d)$ at random, with the constraint to achieve $\Phi_a > 0$ for all $a$ (we sample only the pairs that can contribute to increasing the zeros entries of this vector) and then we randomly pick other pairs until the fixed size of $\X$ is reached. We empirically choose $|\X| = 8$ because it is a good trade-off between two competing effects: if $|\X|$ is too low there is a possibility of over-representing households with elderly members that are fewer and hence more valuable to get the condition $\Phi_a > 0$ for all $a$; on the other hand, very large values of $|\X|$ will tend towards an ``averaging'' effect that leads all \emph{villages} to be very similar to one-another.

\subsubsection*{Model calibration}

Given the samples of villages we now compute the value $\bm{u}$ as the result of the following optimization problem

\begin{align*}
\bm{u} = \underset{\bm{v}~:~\bm{v}^T\bm{1} = {\rm const}}{\rm arg~min}~d_{\rm C}\left(C, T \circ \bm{v}\bm{v}^T\right),
\end{align*}

\LD{where $\left[T \circ (\bm{v}\bm{v}^T)\right]_{ab} = T_{ab} v_av_b$ and $d_{\rm C}$ is a modified Canberra distance}. Let $A, B$ be two symmetric matrices of size $n_{\rm age}$, then

\begin{align*}
d_{\rm C}(A, B) = \sum_{i = 1}^{n_{\rm age}}\sum_{j\leq i}\frac{|\tilde{A}_{ij}-\tilde{B}_{ij}|}{|\tilde{A}_{ij}| + |\tilde{B}_{ij}|}
\end{align*}

where $\tilde{A}$ is the matrix $A$ divided by its mean \LD{(and equivalently $\tilde{B}$ is $B$ divided by its mean)}. \LD{The distance $d_{\rm C}$ is the Canberra distance computed on the matrices $\tilde{A}, \tilde{B}$, instead of $A, B$, hence we refer to it as \emph{modified Canberra distance}.} This choice of the distance is motivated by the two following points

\begin{enumerate}
	
	\item \LD{The entries of $C$ may differ even by a factor $100$} as shown in Figure~\ref{fig:est}. The cosine similarity is meaningful to quantify the proximity of two matrices but it naturally tends to give more weight to entries with a larger magnitude. \LD{For this reason it is unsuited for an optimization as it would poorly estimate the small entries of $C$. On the opposite, the relative distance $d_{\rm C}$ gives approximately the same weight to all matrix entries and can be used for this purpose.}
	
	\item The modified Canberra distance compares a normalized version of the contact matrices because \LD{we are interested in determining them up to a constant factor}. We then have for any $\alpha, \beta >0$, $d_{\rm C}(\alpha A, \beta B) = d_{\rm C}(A, B)$.
	
\end{enumerate}

\subsubsection*{Occupation parameter}

We here detail the strategy to determine the vectors $\bm{y}, \bm{\eta}$ appearing in Figure~\ref{fig:res}b, c, referred to as \emph{occupation} and \emph{compliance} vector respectively.

\medskip

In the PHIRST data collection process, the participants were asked to specify locations or activities in which they spend more than three hours a day for more than three days per week. The options to choose from included: school, university, work, pub, social clubs, hanging out with friends, street vendors and church. We then define a Boolean variable for each person indicating whether or not he/she has a major activity outside the household, \ie if he/she answered positively to \emph{any} of the questions above. The value of $y_a$ is the average of the Boolean indicator for all people of age $a$ in $\X$.

\end{document}